\documentstyle[12pt,epsfig]{article}
\def\beq{\begin{equation}}
\def\eeq{\end{equation}}
\def\bea{\begin{eqnarray}}
\def\eea{\end{eqnarray}}
\def\non{\nonumber}
\def\noi{\noindent}
\def\bib{\bibitem}

\begin{document}

\begin{center} 
{\large \bf \sf 
Construction of some special subsequences within a Farey sequence}

\vspace{1.3cm}

{\sf B. Basu-Mallick$^1$\footnote{e-mail address: biru@theory.saha.ernet.in},
Tanaya Bhattacharyya$^1$\footnote{e-mail address: tanaya@theory.saha.ernet.in}
and Diptiman Sen$^2$\footnote{e-mail address: diptiman@cts.iisc.ernet.in}}

\bigskip

{\em $^1$Theory Group, Saha Institute of Nuclear Physics, \\
1/AF Bidhan Nagar, Kolkata 700 064, India} 

\bigskip

{\em $^2$Centre for Theoretical Studies, Indian Institute of Science, \\
Bangalore 560012, India}
\end{center}

\bigskip
\bigskip

\noi {\bf Abstract}

Recently it has been found that some special subsequences within a Farey
sequence play a crucial role in determining
the ranges of coupling constant for which quantum soliton states can exist 
for an integrable derivative nonlinear Schr\"odinger model. In this article, 
we find a novel mapping which connects two such subsequences belonging to 
Farey sequences of different orders. By using this mapping, we construct an 
algorithm to generate all of these special subsequences 
within a Farey sequence. We also derive the continued fraction expansions 
for all the elements belonging to a subsequence and observe a close connection
amongst the corresponding expansion coefficients. 

\bigskip


\vspace {.1 cm}

\newpage

\noi \section{Introduction}
\renewcommand{\theequation}{1.{\arabic{equation}}}
\setcounter{equation}{0}

\medskip

The concept of Farey sequences appearing in number theory
has recently found interesting applications 
in diverse subjects in physics like phase transitions in one dimensional
statistical models with long-range interaction \cite{kleb,cont,fial}, 
fractal statistics \cite{cruz}, diffraction patterns of aperiodic crystals 
\cite {radu}, 2+1 dimensional gravity \cite {khol} and quantum soliton states 
of an integrable 1+1 dimensional derivative nonlinear Schr\"odinger (DNLS) 
model \cite{basu}. For any positive integer $N$, the Farey sequence
of order $N$ (denoted by $F_N$) is defined to be the set of all the 
fractions $a/b$ in increasing order such that (i) $0 \le a \le b \le N$,
and (ii) $a$ and $b$ are relatively prime (i.e., the greatest 
common divisor of $a$ and $b$ is $1$) \cite{hard,nive}. 
The Farey sequences for the first few integers are given by
\bea
F_1: & & \quad \frac{0}{1} ~~~~ \frac{1}{1} \non \\
F_2: & & \quad \frac{0}{1} ~~~~ {\bf \frac{1}{2}} ~~~~\frac{1}{1} \non \\
F_3: & & \quad \frac{0}{1} ~~~~{\bf \frac{1}{3}} ~~~~\frac{1}{2} ~~~~
{\bf \frac{2}{3}} ~~~~ \frac{1}{1} \non \\
F_4: & & \quad \frac{0}{1} ~~~~{\bf \frac{1}{4}} ~~~~\frac{1}{3} ~~~~
\frac{1}{2} ~~~~
\frac{2}{3} ~~~~{\bf \frac{3}{4}} ~~~~\frac{1}{1} \non \\
F_5: & & \quad \frac{0}{1} ~~~~{\bf \frac{1}{5}} ~~~~\frac{1}{4} ~~~~
\frac{1}{3} ~~~~
{\bf \frac{2}{5}} ~~~~\frac{1}{2} ~~~~{\bf \frac{3}{5}} ~~~~\frac{2}{3} ~~~~
\frac{3}{4} ~~~~{\bf \frac{4}{5}} ~~~~\frac{1}{1} ~.
\label{a1}
\eea
These sequences enjoy many interesting properties \cite{hard,nive}, 
of which we list the relevant ones below.

\noi (i) If $a/b < a'/b'$ are two successive fractions in a Farey 
sequence $F_N$, then 
\beq
a' b ~-~ a b' ~=~ 1 ~, ~~~~ b ~, ~b' ~\le ~N ~~ {\rm and}~~~~ b+b' >N ~.
\label{a2}
\eeq
In this case, it follows that both $a$ and $b'$ are relatively prime to
$a'$ and $b$.

\noi (ii) If $a/b < a'/b' < a''/b''$ are three successive fractions in a 
Farey sequence $F_N$, then 
\beq
\frac{a'}{b'} ~=~ \frac{a+a''}{b+b''} ~.
\label{a3}
\eeq

\noi (iii) For $N \ge 2$, let $n/N$ be a fraction appearing somewhere in the 
sequence $F_N$ (such fractions are denoted by bold letters in Eq. (\ref{a1}) ).
Then the fractions $a_1/b_1$ and $a_2/b_2$ appearing
immediately to the left and to the right respectively of $n/N$ satisfy
\bea
a_1 ~,~ a_2 ~\le ~n ~, \quad {\rm and} \quad a_1 ~+~ a_2 ~=~ n ~, \non \\
b_1 ~,~ b_2 ~<~ N ~, \quad {\rm and} \quad b_1 ~+~ b_2 ~=~ N ~.
\label{a4}
\eea

We have recently investigated the ranges of coupling 
constant (called bands) within which localized $N$-body soliton states 
can be constructed for a quantum integrable DNLS model \cite{basu}. 
Interestingly, it is found that such bands have a one-to-one correspondence 
with the fractions $n/N$, which appear within the sequence $F_N$. 
Consequently, we use the notation $B_{n,N}$ to denote such a band. 
Farey sequences also play a crucial role in finding the end points 
of these bands beyond which localized $N$-soliton states do not exist 
for the DNLS model. It is found that the fractions $a_1/b_1$ and $a_2/b_2$, 
appearing immediately to the left and to the right respectively of $n/N$ 
within the Farey sequence $F_N$, determine the end points of the band 
$B_{n,N}$. Therefore, special subsequences like
$a_1/b_1, n/N, a_2/b_2$ belonging to $F_N$
play a key role in the context of quantum soliton states of DNLS model. 

In this article, we focus our attention on such subsequences within Farey 
sequence and find a novel connection between these subsequences. 
Let us denote the subsequence of three successive fractions,
$a_1/b_1, n/N, a_2/b_2$ in $F_N$, as $F_{n,N}$.
It may be noted that the first subsequence in $F_N$, i.e., $F_{1,N}$,
has a very simple form for all values of $N$. Namely, $F_{1,N}$ consists
of $0/1, 1/N, 1/(N-1)$. However, higher subsequences in $F_N$ (i.e., 
$F_{n,N}$ with $n>1$) cannot be expressed in such a simple way. 
In Sec. 2 of this article, we find a mapping between two subsequences 
within two Farey sequences of different orders. Due to this mapping, 
any higher subsequence in $F_N$ can be generated through 
the first subsequence in some $F_M$, where $M<N$. In this way, 
we find an algorithm to generate all $F_{n,N}$ with $n>1$. In Sec. 3 we 
show that the above mentioned algorithm to generate the elements of $F_{n,N}$
can be expressed in an elegant way through continued fractions. In this 
section we also discuss a method for finding out the successive fraction of 
any given fraction within a Farey sequence. Sec. 4 is the concluding section.

\vspace{1cm}

\noi \section{Mapping between subsequences of Farey sequences}
\renewcommand{\theequation}{2.{\arabic{equation}}}
\setcounter{equation}{0}

\medskip
Following the standard convention \cite{hard}, we call $a/b$ 
an irreducible fraction when $a$ and $b$ are relative prime numbers. 
For any two irreducible positive fractions $a/b$ and $c/d$, we define
\beq
\Delta(a/b, c/d) \equiv cb - ad.
\label{b1}
\eeq
For the case $a/b < c/d$, $ \Delta(a/b, c/d) $ is a positive integer. Let us 
now consider the following theorem which may be regarded as 
the converse of relation (\ref{a2}).

\noi {\bf Theorem 2.1} \quad {\it If $0 \leq a/b < a'/b' \leq 1$ 
are two irreducible fractions which satisfy 
$\Delta(a/b, a'/b') =1$, then they will appear as successive fractions in 
$F_N$ for any value of $N$ which lies in the range $max(b,b')\leq N< b+b'$.} 

\noi {\it Proof} \quad It is obvious that $a/b$ and $a'/b'$ are some 
fractions belonging to $F_N$. Let us now suppose that $a/b$ and $a'/b'$ are 
not successive fractions in $F_N$. Let $c/d$ be a fraction which lies between 
them in $F_N$ (there may be more than one such fraction; in that case, we 
choose any one of them). Then we have two relations like $ \Delta(a/b,c/d)= 
n_1$, and $\Delta(c/d,a'/b')= n_2$, where $n_1, n_2 \ge 1$. Let us multiply 
the first relation by $b'$, the second by $b$, and add these two relations.
Thus we get $d ~\Delta(a/b, a'/b') = b' n_1 + b n_2.$
Since $ \Delta(a/b, a'/b') = 1$, we see that $d = b' n_1 + b n_2$.
However, since $b + b' > N$, and $n_1, n_2 \ge 1$, we necessarily have 
$d = b' n_1 + b n_2 > N$. This proves by contradiction that $a/b$ and $a'/b'$
must be successive fractions in $F_N$.

Let us now consider another theorem which provides a novel mapping 
between two subsequences of Farey sequences with different orders. 

\noi {\bf Theorem 2.2} \quad {\it If $a_1/b_1, a_2/b_2, a_3/b_3$ 
is a subsequence in $F_{b_2}$, then $b_3/(\rho b_3 + a_3), 
b_2/(\rho b_2 + a_2), b_1/(\rho b_1 + a_1)$ is a subsequence in 
$F_{\rho b_2 + a_2}$, where $\rho$ is any positive integer.}

{\it Proof} \quad Since $a_1/b_1, a_2/b_2, a_3/b_3$ 
is a subsequence in $F_{b_2}$, the pairs $(a_i,b_i)$ are relative prime. 
By using this property, it is easy to show that the pairs $(b_i, \rho b_i+a_i)$
are also relative prime. Thus irreducible fractions associated with the pairs 
$(b_i, \rho b_i+a_i)$ are fit candidates to form a subsequence in 
$F_{\rho b_2 + a_2}$. Using the relation (\ref{b1}), we find that
\beq
\Delta\left( \frac{b_2}{\rho b_2 + a_2} ,\frac{b_1}{\rho b_1+a_1}\right) =
\Delta\left(\frac{a_1}{b_1},\frac{a_2}{b_2}\right) ~, ~~
\Delta\left( \frac{b_3}{\rho b_3 + a_3} ,\frac{b_2}{\rho b_2+a_2} \right) =
\Delta\left(\frac{a_2}{b_2},\frac{a_3}{b_3}\right) ~.
\label{b2}
\eeq
Since $a_1/b_1, a_2/b_2, a_3/b_3$ is a subsequence in $F_{b_2}$, 
$\Delta \left( a_1/b_1 , a_2 /b_2 \right) = 
\Delta\left(a_2 /b_2,a_3 /b_3\right) = 1.$ So, from Eq. (\ref{b2}), we obtain
\beq
\Delta\left( \frac{b_2}{\rho b_2 + a_2} ,\frac{b_1}{\rho b_1+a_1}\right) =
\Delta\left( \frac{b_3}{\rho b_3 + a_3} ,\frac{b_2}{\rho b_2+a_2} \right) =1~.
\label{b3}
\eeq
Moreover, using property (\ref{a4}) for the subsequence
$a_1/b_1, a_2/b_2, a_3/b_3$, one finds that
$a_2 = a_1 + a_3$ and $b_2 = b_1 + b_3$. Due to these relations,
it follows that $\rho b_2 + a_2 = (\rho b_1 + a_1) + (\rho b_3 + a_3)$ and 
\beq
max (\rho b_2 + a_2, \rho b_1 + a_1) =
max (\rho b_2 + a_2, \rho b_3 + a_3) = \rho b_2 + a_2 ~.
\label {b7}
\eeq
Consequently, by applying Theorem 2.1 along with relations (\ref{b3}) and 
(\ref{b7}), we find that $b_3/(\rho b_3 + a_3), b_2/(\rho b_2 + a_2), 
b_1/(\rho b_1 + a_1)$ is a subsequence within $F_{\rho b_2 + a_2}$. The 
converse of Theorem 2.2 is also valid and may be proved in a similar fashion.

Starting from any known subsequence in $F_N$, and applying Theorem 2.2 
repeatedly, one can easily construct many other subsequences in $F_{N_1}$, 
where $N_1>N$. We have already noted that $F_{1,N}$ has a very simple form, 
namely, $ 0/1, 1/N, 1/(N-1)$. Since such subsequences cannot be 
generated from some other subsequences by using Theorem 2.2, they
may be classified as `fundamental subsequences'. In the following,
we shall show that all $F_{n_1,N_1}$ with $n_1>1$ are `composite
subsequences', which can be constructed from the fundamental subsequences
by using Theorem 2.2. To construct $F_{n_1,N_1}$, it is only necessary 
to find the first and last fraction appearing within this composite 
subsequence. We can always map the middle fraction of $F_{n_1,N_1}$
to the middle fraction of some fundamental subsequence through 
successive steps given by
\beq
\frac{n_1}{N_1} \stackrel{\rho_1}{\longrightarrow}
\frac{n_2}{N_2} 
\cdots \cdots \cdots \frac{n_i}{N_i} 
\stackrel{\rho_i}{\longrightarrow} \frac{n_{i+1}}{N_{i+1}} 
\cdots \cdots \cdots \frac{n_k}{N_k} 
\stackrel{\rho_k}{\longrightarrow} \frac{1}{N_{k+1}} ~,
\label{b4}
\eeq 
where $\rho_i = \left[ \frac{N_i}{n_i} \right] $
(here the symbol $[q]$ denotes the integer part of $q$),
$n_{i+1}=N_i - \rho_i n_i$, $N_{i+1}=n_i$, and it is assumed that 
$n_{k+1}=1$. It may be noted that, within the fundamental subsequence
$F_{1,N_{k+1}}$, $0/1$ and $1/(N_{k+1}-1)$ appear in the left and the right 
side of the fraction $1/N_{k+1}$ respectively. Applying the mappings of Eq. 
(\ref{b4}) in reverse order to the fractions $0/1$ and $1/(N_{k+1}-1)$,
we obtain 
\bea
\frac{0}{1} &\equiv &\frac{n'_{k+1}}{N'_{k+1}}
\stackrel{\rho_k}{\longrightarrow} \frac{n'_k}{N'_k} \cdots \cdots \cdots 
\frac{n'_{i+1}}{N'_{i+1}} \stackrel{\rho_i}{\longrightarrow} 
\frac{n'_i}{N'_i} \cdots \cdots \cdots \frac{n'_2}{N'_2} 
\stackrel{\rho_1}{\longrightarrow} \frac{n'_1}{N'_1} ~, \non \\
\frac{1}{N_{k+1}-1} & \equiv & \frac{n''_{k+1}}{N''_{k+1}}
\stackrel{\rho_k}{\longrightarrow}
\frac{n''_k}{N''_k} \cdots \cdots \cdots 
\frac{n''_{i+1}}{N''_{i+1}} \stackrel{\rho_i}{\longrightarrow} 
\frac{n''_i}{N''_i} \cdots \cdots \cdots \frac{n''_2}{N''_2} 
\stackrel{\rho_1}{\longrightarrow} \frac{n''_1}{N''_1} ~,
\label{b5}
\eea 
where $n'_i=N'_{i+1}, N'_i= n'_{i+1} + \rho_i N'_{i+1}$ and 
$n''_i=N''_{i+1}, N''_i= n''_{i+1} + \rho_i N''_{i+1}$. Since 
$n'_{k+1} /N'_{k+1}, n_{k+1}/N_{k+1},n''_{k+1}/N''_{k+1}$
represents the fundamental subsequence in $F_{N_{k+1}}$, by applying 
Theorem 2.2 with $\rho=\rho_k$ one can easily show that
$n''_k /N''_k, n_k/N_k, n'_k/N'_k$ is a subsequence
in $F_{N_k}$. Applying Theorem 2.2 repeatedly in this fashion, we find that
\beq
\frac{n'_1}{N'_1}, \frac{n_1}{N_1}, \frac{n''_1}{N''_1}, ~~~~{\rm or,}~~~~
\frac{n''_1}{N''_1}, \frac{n_1}{N_1}, \frac{n'_1}{N'_1},
\label{b6}
\eeq
generates the subsequence $F_{n_1,N_1}$ within $F_{N_1}$, 
when $k$ is an even or odd integer respectively.

Thus we observe that all higher subsequences within $F_{N_1}$ are in fact 
composite subsequences, which can be generated in the above mentioned way.
As an example, let us try to find the composite subsequence $F_{5,39}$ 
which lies within $F_{39}$. By using (\ref{b4}), we find that 
$$ \frac{5}{39} ~ \stackrel{\rho_1=7~~}{\longrightarrow} 
\frac{4}{5} ~ \stackrel{\rho_2=1~~}{\longrightarrow} \frac{1}{4}~.$$
Thus $1/4$ is the middle fraction of the corresponding 
fundamental subsequence $F_{1,4}$. The first and last 
fraction of this fundamental subsequence are given by $0/1$ and $1/3$ 
respectively. Applying the mapping (\ref{b5}) to these fractions, we obtain
\bea
&&\frac{0}{1} ~ \stackrel{\rho_2=1~~}{\longrightarrow} 
\frac{1}{1} ~ \stackrel{\rho_1=7~~}{\longrightarrow} \frac{1}{8} ~,~~\non \\
&&\frac{1}{3} ~ \stackrel{\rho_2=1~~}{\longrightarrow} 
\frac{3}{4} ~ \stackrel{\rho_1=7~~}{\longrightarrow} \frac{4}{31} ~. \non
\eea
Since this is two step process ($k=2$), by using (\ref{b6})
we find that $ F_{5,39}= 1/8, 5/39,4/31$.

\medskip
\noi \section{Farey subsequences and continued fractions}
\renewcommand{\theequation}{3.{\arabic{equation}}}
\setcounter{equation}{0}
\medskip

The above mentioned procedure of generating a subsequence of Farey sequence 
can be described in an elegant way with the help of continued fractions. 
We first discuss the idea of a continued fraction \cite{nive}. Any positive 
real number $x$ has a simple continued fraction expansion of the form
\beq
x ~=~ n_0 ~+~ \frac{1}{n_1 ~+~ \frac{1}{n_2 ~+~ \frac{1}{n_3 ~+~ \cdots}}} ~,
\label{c1}
\eeq
where the $n_i$'s are integers satisfying $n_0 \ge 0$, and $n_i \ge 1$ for 
$i \ge 1$. Given a number $x$, the integers $n_i$ can be found as follows. 
Let us first define $x_0 = x$ and $n_0 = [x_0]$.
We then recursively define $x_{i+1} = 1/(x_i - n_i)$,
and obtain $n_{i+1} = [x_{i+1}]$ for $i =0,1,2,\cdots$. This
expansion ends at a finite stage with a last integer $n_k$ if $x$ 
is rational. In that case, we can assume that the last integer satisfies $n_k 
\ge 2$ (if $n_k$ is equal to $1$, we can stop at the previous stage and 
increase $n_{k-1}$ by 1). With this convention for $n_k$, the continued 
fraction expansion written in the form 
$$x = [n_0,n_1,n_2,\cdots,n_{k-1},n_k] ~,$$
is unique for any rational number $x$. If $x$ is an irrational number, the 
continued fraction expansion does not end, and it is unique.

As shown in Eq. (\ref{b6}), the subsequence $F_{n_1,N_1}$ contains 
the fractions $n'_1/N'_1$, $n_1/N_1$, and $n''_1/N''_1$. 
In Sec. 2 we have given an algorithm for finding out the fractions 
$n'_1/N'_1$ and $n''_1/N''_1$ around the known middle fraction 
$n_1/N_1$. Let us now try to find the continued fraction expansion
for all of these elements in $F_{n_1,N_1}$. To this end, 
we observe that the relation between fractions $n_i/N_i$ and 
$n_{i+1}/N_{i+1}$ appearing in Eq. (\ref{b4}) can be written in the form 
\beq
\frac{n_i}{N_i} = \frac{1}{\rho_i + \frac{n_{i+1}}{N_{i+1}}} ~.
\label{c2}
\eeq
Since $n_{i+1}/N_{i+1}$ is the middle term of a composite subsequence having
value less than $1$, the continued fraction expansion for $n_{i+1}/N_{i+1}$ 
can be written as $[0,X]$, where $X \equiv n_1, n_2, \cdots n_k$. 
Then, due to Eq. (\ref{c2}), it follows that 
\beq
\frac{n_{i+1}}{N_{i+1}} = [0,X] 
\Longrightarrow \frac{n_i}{N_i} = [0, \rho_i , X] ~.
\label{c3}
\eeq
Taking $n_{k+1}/N_{k+1} \equiv 1/N_{k+1} = [0,N_{k+1}]$ as the `initial
condition', and applying Eq. (\ref{c3}) repeatedly, we easily obtain 
\beq
\frac{n_1}{N_1} = [0, \rho_1 , \rho_2, \cdots , \rho_k , N_{k+1}] ~.
\label{c4}
\eeq
It may be noted that the fractions $n'_i/N'_i$ and 
$n'_{i+1}/N'_{i+1}$ (or, $n''_i/N''_i$ and $n''_{i+1}/N''_{i+1}$) appearing in 
Eq. (\ref{b5}) also satisfy relations exactly of the form (\ref{c2}) and
(\ref{c3}). Taking $0/1=[0]$ and $1/(N_{k+1}-1) = [0, N_{k+1}-1]$ as 
initial conditions, and applying relations of the form (\ref{c3}), we derive 
the continued fraction expansion for $n'_1/N'_1$ and $n''_1/N''_1$ as
\bea
&&\frac{n'_1}{N'_1} = [0, \rho_1 , \rho_2, \cdots , \rho_k] ~, \non \\
&&\frac{n''_1}{N''_1} = [0, \rho_1 , \rho_2, \cdots , \rho_k , N_{k+1} -1] ~.
\label{c5}
\eea
Comparison of Eqs. (\ref{c4}) and (\ref{c5}) reveals that 
the continued fraction expansions of three elements within subsequence 
$F_{n_1,N_1}$ are closely connected with each other. Previously we have 
established such a connection through a completely a different route by 
using some properties of the continued fraction expansion \cite{basu}.
By finding the continued fraction expansion of the middle term $n_1/N_1$
and applying Eq. (\ref{c5}), one can easily construct the 
subsequence $F_{n_1,N_1}$. As an explicit example, let us try to construct 
the subsequence $F_{9,25}$. The continued
fraction expansion of its middle term is given by $9/25 = [0,2,1,3,2]$. 
Eq. (\ref{c5}) then gives $n'_1/N'_1 =[0,2,1,3]=4/11$ and 
$n''_1/N''_1=[0,2,1,3,1]=5/14$. Since $k=3$ for this case, 
using Eq. (\ref{b6}) we obtain $F_{9,25} = 5/14, 9/25, 4/11$.

Finally, it may be noted that our procedure of finding a subsequence of Farey 
sequence can also be used to find the successive term of any given fraction 
within a Farey sequence. Let us try to find the successive term 
on the right side of the fraction $a/b$ within a Farey sequence $F_N$.
By using the above mentioned procedure of continued fractions, we can 
easily find such successive term of $a/b$ within the Farey sequence $F_b$. 
Let us denote this term as $a_0/b_0$. 
Since we have $\Delta ( a/b, a_0/b_0) =1$ and $b>b_0$, due to Theorem 2.1 it is 
clear that $a/b $ and $a_0/b_0$ will be successive terms within any $F_N$
with $N$ lying in the range $b \leq N < b+b_0$. Next, we want to find the 
successive term of $a/b$ within $F_N$ for the case $N\geq b+b_0$. It is easy 
to see that $$\Delta \left( \frac{a}{b} , \frac{la+a_0}{lb+b_0} \right) 
= \Delta \left( \frac{a}{b} , \frac{a_0}{b_0} \right) =1 ~,$$
where $l$ is any positive integer. Consequently, due to Theorem 2.1,
$a/b $ and $(la+a_0)/(lb+b_0)$ will be successive terms within any $F_N$
with $N$ lying in the range $lb+b_0 \leq N < (l+1)b+b_0$. Thus, for any 
given value of $N$, the successive term on the right side
of the fraction $a/b$ within the sequence $F_N$ is obtained as
\beq
\frac{a_l}{b_l} = \delta_{l,0} ~\frac{a_0}{b_0} + 
(1-\delta_{l,0}) ~\frac{la+a_0}{lb+b_0} ~, 
\label{c6}
\eeq
where $l = [(N-b_0)/b]$. As a concrete example, let us try to calculate 
the successive term on the right side of $9/25$ within the 
Farey sequence $F_{100}$. By using the method of continued fractions, we 
have already found in the preceding paragraph that $4/11$ appears in the
right side of $9/25$ within the Farey sequence $F_{25}$. Since $a/b=9/25$ and
$a_0/b_0= 4/11$ in this case, we have 
$l=[(100-11)/25]=3$. Eq. (\ref{c6}) then gives the successive 
term of $9/25$ within $F_{100}$ as $a_3/b_3= 31/86$. 

\vspace{1cm}

\noi \section{Concluding remarks}
\medskip

In this work, we have studied some special subsequences within a Farey 
sequence which appear naturally in the context of quantum soliton states 
for an integrable derivative nonlinear Schr\"odinger model. 
In particular, we have found a novel mapping (as stated in Theorem 2.2)
through which one can generate many such subsequences within Farey 
sequences from the knowledge of any given subsequence. 
This mapping allows us to classify all of these subsequences into two
types - the `fundamental subsequences' and the `composite subsequences'. We 
find that all composite subsequences can be constructed by mapping them to 
fundamental subsequences which are always expressed in a known simple form. 
In this way, we obtain an algorithm to generate all composite subsequences 
within a Farey sequence. We have also derived the continued fraction expansions
(\ref{c4}) and (\ref{c5}) for all elements within a subsequence, and have
found a close connection amongst the corresponding expansion coefficients. 
Consequently, our algorithm for generating all subsequences 
within a Farey sequence can be expressed in an elegant way
through the continued fraction expansions.

\end{document}